\documentstyle[12pt]{article}
\newcommand{\beq}{\begin{equation}}
\newcommand{\eeq}{\end{equation}}
\begin{document}
\begin{titlepage}
\begin{flushleft}
       \hfill                      {\tt astro-ph/yymmdd}\\
       \hfill                      UUITP-15/96\\
       \hfill                       July 1996\\
\end{flushleft}
\vspace*{3mm}
\begin{center}
{\LARGE Distortion of the acoustic peaks in the CMBR due to a
primordial magnetic field\\ }
\vspace*{12mm}
{\large Jenni Adams\footnote{E-mail: jenni@teorfys.uu.se. },
Ulf H. Danielsson\footnote{E-mail: ulf@teorfys.uu.se},
Dario Grasso\footnote{EEC fellow,E-mail: grasso@teorfys.uu.se.} }\\
\vspace{1mm}
{\large and } \\
\vspace{1mm}
{\large H\'ector Rubinstein\footnote{E-mail: rub@physto.se}} \\
\vspace{4mm}
{\em Institutionen f\"{o}r Teoretisk Fysik \\
Box 803\\
S-751 08  Uppsala \\
Sweden \/}\\

\vspace*{10mm}
\end{center}

\begin{abstract}
In this paper we study the effect of a magnetic field on the 
fluctuation spectrum of the cosmic microwave background. We find that 
upcoming measurements might give interesting bounds on large scale 
magnetic fields in the early Universe. If the effects are seen, it might 
be possible to establish the presence of different fields in different 
patches of the sky. Absence of any effect, will provide by one order 
of magnitude a better limit for a primordial field,
now given by nucleosynthesis. 
\end{abstract}

\end{titlepage}

\section{Introduction}

Very little is known about cosmic magnetic fields, both
those that exist today and those
in the early Universe. 
Even the stability of large fields is open to 
conjecture\cite{stability}. In the galaxy one measures a field of the 
order of $10^{-6}$ Gauss but its origin remains a 
mystery \cite{origin}. If it is primordial, it could have resulted 
from a compression of a 
cosmological field corresponding to around $10^{-9}$ Gauss today. 
This is comparable to limits set for fields on the horizon scale 
using Faraday 
rotation on faraway galaxies and quasars. When traced back 
in time such a field becomes quite strong since $B \sim 1/a^2$ 
where $a$ is the scale factor. 

The presence of primordial fields is a hotly debated issue. For a 
long time the dynamo mechanism with small seed fields was favored, 
but the recent discovery of damped Ly$\alpha$ lines in QSO's 
indicates that primordial fields existed at early times. Moreover,
there are problems with the dynamo mechanism. For a short discussion 
and further references see \cite{Loeb}. 

The QSO measurements are consistent with
having $\mu$G fields at $z^{abs}=2$. 
It is not unreasonable to expect that such fields might have had 
measurable effects on physics in the early Universe.
One such possibility was studied in \cite{Grasso} where it was found 
that nucleosynthesis bounded the field to $10^{11}$ - $10^{12}$ 
Gauss (lower limit for fields homogeneous on the horizon scale) at a 
time when $T=10^9K$.
This corresponds to between $10^{-6}$ and $10^{-7}$ Gauss today.
Another way to set limits, this time at last scattering, is to study
Faraday rotation directly in the CMB. In \cite{Loeb} it is claimed 
that it should be possible to reach a field equivalent to 
$10^{-9}$ Gauss today in this way. Existence of these fields may also
have a large impact on structure formation \cite{Olinto}.

In this paper we will discuss the possibility of taking advantage of
the many upcoming precision measurements of CMB anisotropies. These
measurements, involving satellites, ground interferometry,
and balloons\cite{Tegmark},
promise to provide us with accurate values of many cosmological 
parameters.

When primordial density fluctuations, perhaps generated by inflation, 
enter the horizon some time before last scattering, they initiate 
acoustic
oscillations in the plasma. These oscillations distort the primordial
spectrum of fluctuations and their effect can be studied today. 
Clearly the result will be very sensitive to the physics of the 
plasma and this is
the reason for the present optimism.

As we will argue in this paper, magnetic fields of reasonable 
magnitude will also affect the plasma leaving a possibly measurable 
imprint on the CMB.
There are several exciting possibilities that may be detectable:
a) different types of waves (see below) depending on the 
properties of the primordial fluid creating different displacements 
of acoustic peaks and
changing their magnitudes, b) anisotropies 
(at the level of $10^{-6}$) 
that maybe different in different areas of the sky, 
signaling the presence of
magnetic field patches in the early Universe.

The paper is organized as follows. In section 2 we recall some 
elementary magnetohydrodynamics describing 
the kind of waves which might be occurring 
in the
plasma. In section 3 we discuss the various types in detail 
and give some qualitative and quantitative predictions 
on how they might affect the
CMB. Section 4 contains our conclusions.

\section{Some magnetohydrodynamics}

A rigourous analysis of the effects of the magnetohydrodynamics 
modes on the 
CMB requires the introduction of a multifluid theory and a general
relativistic treatment. However a brief description of the main 
features of the magnetohydrodynamics of a nonrelativistic one 
component charged fluid 
is physically illuminating and will occupy this section. 

We will be considering a magnetic field homogenous on scales larger
than the scale of plasma oscillations. We will therefore assume a
background magnetic field 
${\bf B} _0$ constant in space. 
The actual field is $ 
{\bf B}_0 + {\bf B} _1$ where ${\bf B} _1$ is a small perturbation.
We assume that the electric conductivity of the medium is 
infinite, thus the magnetic flux is constant in time.
Then, due to the expansion of the Universe, ${\bf B}_0 
\propto a^{-2}$.    
We neglect here any dissipative effect, due for example to
a finite viscosity and heat conductivity \cite{Olinto}. In other 
words we are assuming that $\lambda = \frac{2\pi}{k} \gg 
l_{diss}$. This is justified for the large scale fields that we are
considering.

Within these assumptions the linearized equations of MHD in comoving 
coordinates are: 

\begin{equation}
  \dot \delta + {{\bf \nabla \cdot v_{1}}\over a} = 0   ,
        \label{den}
\end{equation}

\begin{equation}
   {\bf \dot{v}}_{1} + {\dot{a}\over a}{\bf v}_{1} + 
   {c_{S}^{2}\over a}
   {\bf \nabla} \delta + {{\bf \nabla} \phi_{1}\over a} +
   {{\bf \hat{B}}_{0} \times \left(\dot{\bf v}_{1} \times 
   {\bf \hat{B}}_{0}\right)
   \over 4\pi a^{4}} +
   {{\bf \hat{B}}_{0} \times \left({\bf \nabla} \times 
   {\bf \hat{B}}_{1}\right)
   \over 4\pi\rho_{0} a^{5}} =0 ,
        \label{vel}
\end{equation}

\begin{equation}
        \partial_{t}{\bf \hat{B}}_{1} = 
        {{ \bf \nabla} \times \left(\bf{v}_{1} \times 
        {\bf \hat{B}}_{0}\right)
        \over a}  ,
        \label{b1}
\end{equation}

\begin{equation} 
{\bf \nabla}^{2} \phi_{1} = 4\pi G \rho_{0} \left( \delta + 
{{\bf \hat{B}}_{0}\cdot {\bf \hat{B}}_{1} \over 4\pi \rho_{0} a^{4}} 
\right)  
    \label{phi}
\end{equation}
and
\begin{equation}
        {\bf \nabla \cdot \hat{B}}_{1} = 0      ,
        \label{divergenceless}
\end{equation}
where ${\bf \hat{B}} \equiv {\bf B}a^{2}$ and 
$\delta=\frac{\rho_1 }{\rho_0}$, $\phi_{1}$ and $v_{1}$ are 
small perturbations 
on the background density, gravitational potential and velocity
respectively. $c_{S}$ is the sound velocity. 
Neglecting its direct gravitational influence, the magnetic field
couples to fluid dynamics only through the last two terms in 
eq.\ref{vel}.
The first of these terms is due to the displacement current
contribution to ${\bf \nabla} \times {\bf B}$ whereas the latter 
account for the magnetic force of the current density.
The displacement current term can be neglected provided that 
$ v_{A} = B_{0}/\sqrt{4\pi\rho} \ll c_{S}$, where $v_{A}$ is the
Alfv\'en velocity.

Let us now discuss the basic properties of the solutions of
these equations, ignoring for the moment the
expansion of the Universe.\footnote{The full solutions are given
in \cite{Olinto}.} 
A useful reference on this subject is \cite{mhd}.

Without a magnetic field there is only the ordinary sound wave 
involving density fluctuations
and longitudinal velocity fluctuations (i.e. along the wave vector). 
In the
presence of a magnetic field, however, there are no less than three 
different waves:

{\it 1. Fast magnetosonic waves.}

In the limit of small magnetic fields these waves become the ordinary 
sound waves. Their velocity,
$c_+$, is given by
\begin{equation}
c_+^2 \sim c_S^2 + v_A^2 \sin ^2 \theta , \label{c+}
\end{equation}
where $\theta$ is the angle between $\bf k$ and ${\bf B}_0$.
Fast magnetosonic waves involve fluctuations in the velocity,
density, magnetic field and gravitational field. The velocity 
and density fluctuations are out-of-phase by $\pi/2$. 
Equation (\ref{c+}) is valid for
$v_A << c_S$. For such fields the wave is approximatively 
longitudinal.

{\it 2. Slow magnetosonic waves.}

Like the fast waves, the slow waves  involve both density and 
velocity fluctuations. The
velocity is however fluctuating both longitudinally and 
transversely even
for small fields. The velocity of the slow waves is approximatively
\begin{equation}
c_-^2 \sim v_A^2 \cos ^2 \theta   .
\end{equation}

{\it 3. Alfv\'en waves}

For this kind of waves ${\bf B}_1$ and $\bf v_{1}$ lie in a plane 
perpendicular to the plane through $\bf k$ and ${\bf B}_0$.
In contrast
to the magnetosonic waves, the Alfv\'en waves are purely 
rotational, thus they involve no density fluctuations. 
Alfv\'en waves are linearly polarized. 
Their velocity of propagation is
\begin{equation}
c_A^2 = v_A^2 \cos ^2 \theta    .
\end{equation}

One should note that
for $v_A$ comparable to  both $c_S$ and the speed of light, 
the formula for the velocity of the Alfv\'en waves remains 
uncorrected
while the velocity of the magnetosonic waves are given by
\begin{equation}
c_\pm ^2 = \frac{c_S^2  (1+v_A^2 \cos ^2 \theta /c^2 ) 
+ v_A^2 \pm ((c_S^2  (1+v_A^2 \cos ^2 \theta /c^2 ) 
- v_A^2)^2 + 4 v_A^2 c_S^2 \sin ^2 \theta /c^2 )^{1/2})}
{2(1+v_A^2/c^2)}   .
\end{equation}

\section{Effects on the CMB}

The fluctuations in the CMB can be divided into primary and 
secondary fluctuations. The primary fluctuations involve effects 
coming directly from the density fluctuations and also from 
Doppler shifts from velocity
fluctuations and gravitational redshifts.

We will concentrate on these primary effects and show that the 
presence of a magnetic field will change the predicted spectrum of 
fluctuations by changing the speed of sound.

\subsection{The fast magnetosonic waves}

The simplest, and most important, case is the fast wave. Let us 
consider the equations describing the oscillating baryon and 
photon fluid in conformal Newtonian gauge using conformal time, 
see e.g. \cite{MB} for the case without magnetic field. 
They are (for small $v_A$):
\begin{equation}
\dot{\delta} _b + V_b - 3 \dot{\phi} =0 ,
\end{equation}
\begin{equation}
\dot{V}_b + \frac{\dot{a}}{a} V_b - c_b^2 k^2 \delta _b + k^2 \psi
+\frac{a n_e \sigma _T (V_b - V_{\gamma})}{R}
-\frac{1}{4 \pi \hat{\rho} _b a} {\bf k} \cdot ({\bf \hat{B}} _0
\times ({\bf k} \times {\bf \hat{B}} _1)=0  , \label{tjo}
\end{equation}
\begin{equation}
\dot{\delta}_{\gamma } +\frac{4}{3} V_{\gamma} - 4 \dot{\phi}=0  
\end{equation}
and
\begin{equation}
\dot{V} _{\gamma} - k^2 (\frac{1}{4} \delta _\gamma - 
\sigma _\gamma )
- k^2 \psi - a n_e \sigma _T (V_b - V_ \gamma ) =0 ,
\end{equation}
where $V= i {\bf k} \cdot {\bf v}$ and $R= \frac{3 \rho _b}{4 
\rho _\gamma}$. $c_b$ is the baryon sound velocity in the absence 
of interactions with the photon gas.
We have also for convenience
defined $\rho _b = \frac{\hat{\rho}_b}{a^3}$ and ${\bf B} = 
\frac{{\bf \hat{B}}}{a^2}$.
The terms with $\sigma _T$ are due to Thompson scattering and couple
the photons and the baryons. This term can be eliminated between the 
equations. If furthermore tight coupling is assumed (implying e.g.
$V_b \sim V_{\gamma}$), one can derive an equation for the density
fluctuations only. If $c_b \sim 0$ one finds 
that in the absence of magnetic fields
the effective sound velocity is
\begin{equation}
c_S^2 = \frac{1}{3} \frac{1}{1+R} .
\end{equation}
Thus through tight coupling
the photons provide the baryon fluid with a pressure term and a 
non-zero sound velocity arises.

With a magnetic field we need one more equation:
\begin{equation}
{\bf \dot{\hat{B}}} _1 = i ({\bf \hat{B}} _0 \cdot {\bf k}) {\bf v} _b
-i ({\bf k} \cdot {\bf v}_b ) {\bf \hat{B}} _0  .
\end{equation}
Assuming longitudinal waves we find the last term of equation 
(\ref{tjo}) to be
\begin{equation}
- v_A^2 \sin ^2 \theta k^2 \delta _b  ,
\end{equation}
as expected from the previous section.

Hence we find, 
to this order of approximation, 
that the only effect of the magnetic field is a change in the 
speed of sound. 
A simple way to account
for a magnetic field is therefore to change
\begin{equation}
c_b^2 \rightarrow c_b^2 + v_A^2 \sin ^2 \theta    . 
\end{equation}
We have computed the microwave background spectrum 
with this adjustment of the 
sound velocity using the code of \cite{SZ}.

An extra step in the calculation of the CMB anisotropy arises due to
the fact that the velocity of the fast waves 
depends on the angle between the wave-vector and the 
magnetic field.
As mentioned previously we are assuming a magnetic 
field that is varying in
direction on scales larger than the scale of the fluctuation. 
Hence we should sum over
all wave-vectors with the angle between the magnetic field and 
the line of sight fixed. Different patches
of the sky might therefore show different fluctuation 
spectra depending on this angle. In this paper we will 
only be considering an all-sky
average assuming a field that is varying in direction on very 
large scales.
For this reason we also sum 
over the angle between the field 
and the line-of-sight.
In practice, it is easier to reverse the order of the sum
and the calculation of the microwave background anisotropy.
The result of this procedure is shown in figure 1.

We have assumed a magnetic field that gives a maximum increase in 
$c_S^2$ of $0.05 c^2$ at last scattering,
i.e. $v_A^2 \sim 0.05 c^2$.
This corresponds to $2 \times 10^{-7}$ Gauss today. For
a comparison consider figure 2 which shows the effect of a 
20\% decrease
of baryons. Around the first peak the effects are comparable. 
This allows us to obtain a rough estimate for the magnitude 
of the magnetic fields which should be able to be detected
by future measurements of the microwave background anisotropy.
The process of parameter determination using a 
maximum likelihood fit of the observed 
multipole coefficients is discussed in\cite{Tegmark,Jungman}.
Assuming knowledge of the other 
cosmological parameters which affect the
microwave background spectrum, 
a prediction of $\Omega _b$ accurate to
the order of a percent or 
so should be obtainable.
This translates into a limit on the current strength of magnetic
fields which were present in the early Universe, of the order of 
$5 \times 10^{-8}$ Gauss.
 
On very large scales, larger than the characteristic scale 
of the magnetic field, the effect will 
presumably be averaged out and the precise
shape of the curve will depend on this scale. 
The curve in figure 1 is therefore not applicable
for the very lowest values of $l$, 
if we assume a field varying on, say, the horizon scale.

The approximations we have used can only be trusted for large scales, 
that means
late times for the kind of fields we are considering. For earlier 
times the fields are too strong and the Alfv\'en velocity too high.
It is therefore possible that an accurate treatment of the waves 
might turn 
up even more pronounced effects at small scales.

\subsection{The slow waves}

These waves are a little bit more complicated to handle than the fast
ones, even at low magnetic fields  
because the equations do not decouple in a simple way.
The reason is that they involve
both longitudinal and transverse velocity fluctuations.

It is interesting to note, however, that depending on initial 
conditions they should 
be excited with an amplitude fixed relative to the fast waves.
To illustrate this point we will consider a rather naive toy model.
Using the initial conditions $\dot{\delta} (0) =0$ and ${\bf v} =0$ 
we find
(using WKB)
\begin{equation}
\rho \sim \alpha _+ \cos \omega _+ t + \alpha _- \cos \omega _- t
+ \mbox{constant}    ,
\end{equation}
where $\omega _\pm = c_\pm k$. To fix the ratio $\alpha _- / 
\alpha _+$
we need one further initial condition on ${\bf B}_1$. It is 
reasonable to
assume 
\begin{equation}
{\bf B} _{1} (0) =0 ,
\end{equation}
i.e., all fluctuations of the magnetic field (on this scale) are 
due to
fluctuations of the plasma initiated when entering the horizon. 
Using
\cite{mhd} one can show that
\begin{equation}
\alpha _- / \alpha _+ \sim v_A^2 / c_S^2  .
\end{equation}
Since the velocity of the slow waves are much smaller than
the velocity of the fast waves for small fields, we conclude that 
the Doppler peaks should have a long period modulation. 
Further details
will be presented in a 
future publication.

\subsection{Alfv\'en waves}

As discussed in the previous section the Alfv\'en waves are purely
rotational and involve no fluctuations in the density of the 
photon and baryon fluids. 

With initial conditions like the ones above one sees that the 
Alfv\'en 
waves will not be excited. However, one could reverse the reasoning 
and
use these waves to probe the initial conditions. 
They should be well suited for the detection
of turbulent, rotational velocity perturbations
in the early Universe such as those that might be generated from 
primordial phase-transitions. 
Isocurvature initial conditions are probably the most suitable to 
excite the 
Alfv\'en waves. 

The equation describing the waves are 

\begin{equation}
        \delta_{b} = 0  ,
\end{equation}

\begin{equation}
        {\bf \dot{v}}_{b} + \frac{\dot a}{a}{\bf v}_{b} 
        + \frac {a n_e \sigma _T 
        ({\bf v}_b - {\bf v}_{\gamma})}{R} -
        i \frac {({\bf k \cdot \hat{B}_{0}})} {4\pi \hat{\rho}_{b} a} 
        {\bf \hat{B}}_{1} =0  ,
\end{equation}

\begin{equation}
        \delta_{\gamma} = 0 
\end{equation}
and
\begin{equation}
        {\bf \dot{v}}_{\gamma} -a n_e \sigma _T ({\bf v}_b - 
        {\bf v}_{\gamma})
        = 0   .
        \end{equation}
As expected, in this case the photon velocity is only affected 
by the baryon
velocity through Thompson  scattering.

It is evident that Alfv\'en waves give rise only to a Doppler 
effect on the CMB. 
As with the slow waves, we do not present any numerical estimate 
of the effects of the Alfv\'en waves. This will be done in detail 
in a forthcoming 
paper. Here we only wish to point out that since we do not have any
cancellation between Doppler and gravitational effects for this 
kind of 
waves, they could provide a more clear signature of the presence of 
magnetic fields at the last scattering surface.

\section{Conclusions}

In this paper we have taken some preliminary steps towards 
understanding
the effects of magnetic fields on the CMB.

We have found that the limits one can set are comparable, or 
better than
what can be 
achieved by other means, for example nucleosynthesis \cite{Grasso}.
Fields below $10^{-7}/a^2$ Gauss should be accessible in planned
experiments.
The possibility of finding anisotropies in different sectors of 
the sky
and determine their nature is a possibility that is exciting. 
Depending on the
scales this may yield information on the age of these fields 
and their spatial
extent.

We have been considering magnetic fields on scales larger than the
characteristic wavelengths of the acoustic waves. It is also 
important
to investigate the possible effects due to random fields on 
smaller scales.

Clearly it is important to study these possible effects in more 
detail and
thereby take advantage of the upcoming precision measurements
of the cosmic microwave bacground.

\section*{Acknowledgements.}

We wish to thank the participants of the NORDITA Uppsala 
workshop where many
of these ideas took form.
In particular we want to thank Lars Bergstr\"om and 
Anders I Eriksson.

\newpage
\noindent{\bf \huge Figure Captions}
\\
\\
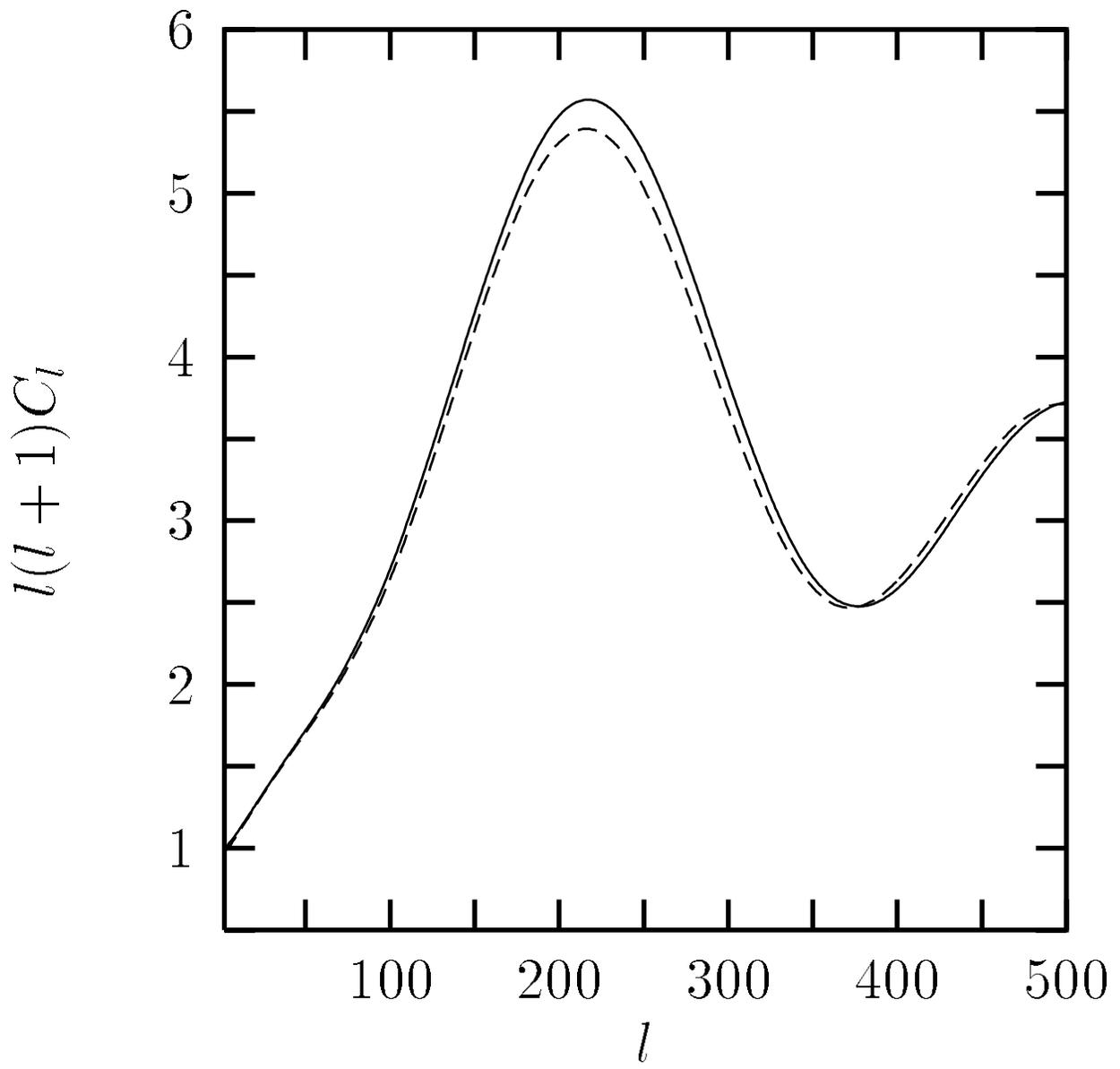
Figure 1:
The effect of a cosmic magnetic field on the multipole
moments. The solid line shows the prediction of a standard CDM cosmology
($\Omega=1$,$h=0.5$, $\Omega_{\rm B}=0.05$) with an $n=1$ primordial
spectrum of adiabatic fluctuations. The dashed line shows the effect 
of adding
a magnetic field equivalent to $2 \times 10^{-7}$ Gauss today.
\\
\\
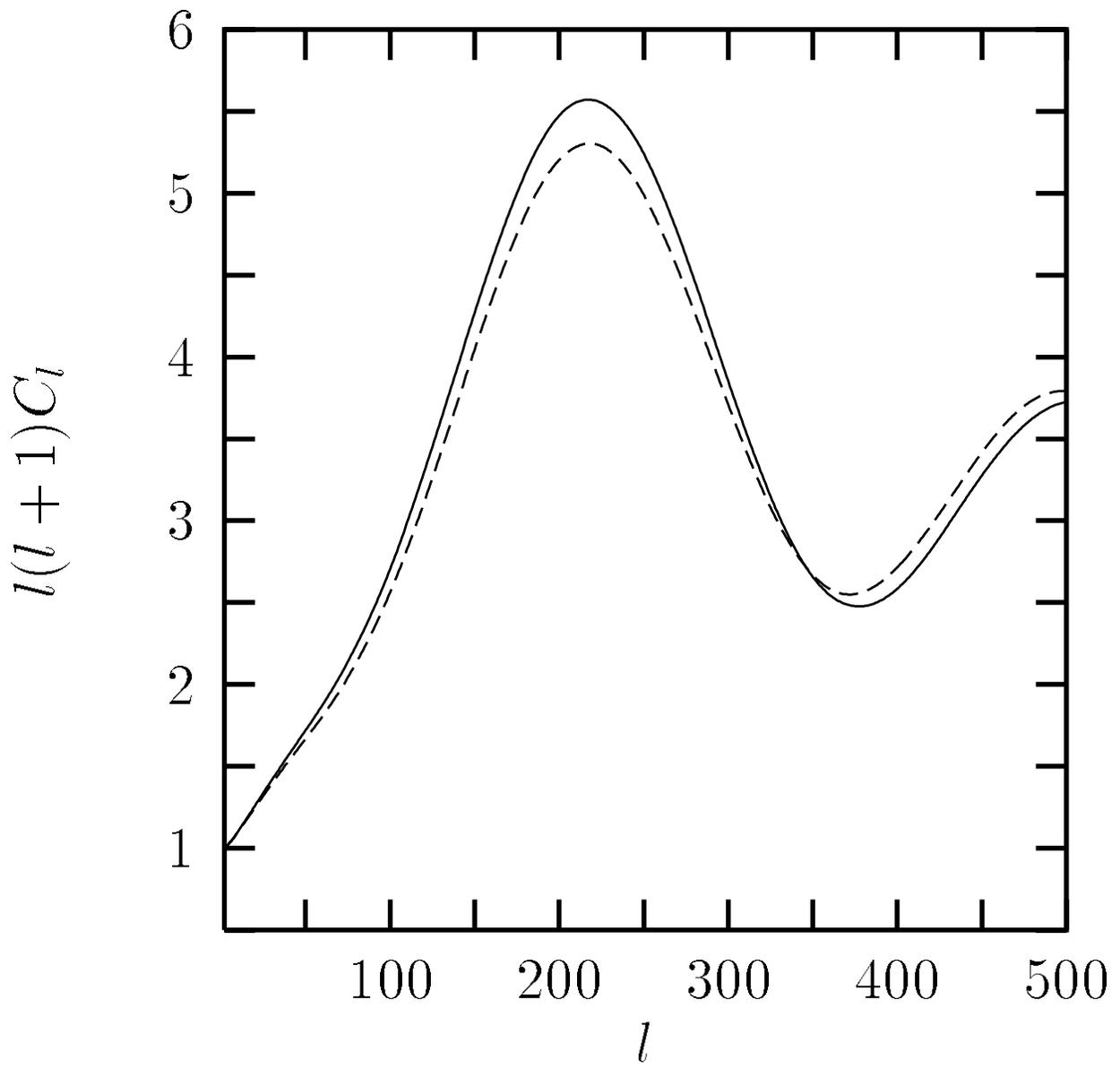
Figure 2: The effect of lowering the baryon fraction by 20 \%
\end{document}